\documentclass[%
 reprint,
 amsmath,amssymb,
 aps,
]{revtex4-2}

\usepackage{graphicx}
\usepackage{dcolumn}
\usepackage{bm}

\begin{document}

\title{\boldmath Radiation Damage of TF-1 and PbWO$_4$ Crystals with 20 MeV Electrons }

\author{Hamlet G.~Mkrtchyan}
\email[Corresponding author, ]{mkrtchyan@yerphi.am}
\affiliation{A.~I. Alikhanyan National Science Laboratory (Yerevan Physice Institute) foundation,\\
02 Alikhanyan Brothers Street, Yerevan 0036, Armenia}

\author{Arthur H.~Mkrtchyan}
\affiliation{A.~I. Alikhanyan National Science Laboratory (Yerevan Physice Institute) foundation,\\
02 Alikhanyan Brothers Street, Yerevan 0036, Armenia}

\author{Vardan H.~Tadevosyan}
\affiliation{A.~I. Alikhanyan National Science Laboratory (Yerevan Physice Institute) foundation,\\
02 Alikhanyan Brothers Street, Yerevan 0036, Armenia}

\author{Hrachya H.~Marukyan}
\affiliation{A.~I. Alikhanyan National Science Laboratory (Yerevan Physice Institute) foundation,\\
02 Alikhanyan Brothers Street, Yerevan 0036, Armenia}

\author{Argine S.~Hakobyan}
\affiliation{A.~I. Alikhanyan National Science Laboratory (Yerevan Physice Institute) foundation,\\
02 Alikhanyan Brothers Street, Yerevan 0036, Armenia}

\author{Ashot S.~Hakobyan}
\affiliation{A.~I. Alikhanyan National Science Laboratory (Yerevan Physice Institute) foundation,\\
02 Alikhanyan Brothers Street, Yerevan 0036, Armenia}

\author{Arthur A.~Hoghmrtsyan}
\affiliation{A.~I. Alikhanyan National Science Laboratory (Yerevan Physice Institute) foundation,\\
02 Alikhanyan Brothers Street, Yerevan 0036, Armenia}

\author{Diana G.~Khurshudyan}
\affiliation{A.~I. Alikhanyan National Science Laboratory (Yerevan Physice Institute) foundation,\\
02 Alikhanyan Brothers Street, Yerevan 0036, Armenia}

\author{Nina N.~Prazyan}
\affiliation{A.~I. Alikhanyan National Science Laboratory (Yerevan Physice Institute) foundation,\\
02 Alikhanyan Brothers Street, Yerevan 0036, Armenia}

\author{Albert H.~Shahinyan}
\affiliation{A.~I. Alikhanyan National Science Laboratory (Yerevan Physice Institute) foundation,\\
02 Alikhanyan Brothers Street, Yerevan 0036, Armenia}

\author{Adelina S.~Stepanyan}
\affiliation{A.~I. Alikhanyan National Science Laboratory (Yerevan Physice Institute) foundation,\\
02 Alikhanyan Brothers Street, Yerevan 0036, Armenia}

\author{Lusine R.~Vahradyan}
 \affiliation{A.~I. Alikhanyan National Science Laboratory (Yerevan Physice Institute) foundation,\\
02 Alikhanyan Brothers Street, Yerevan 0036, Armenia}

\date{\today}

\begin{abstract}
We studied the radiation hardness of two types of crystals, lead glass TF-1 and lead tungstate PbWO$_4$, using a 20 MeV electron beam from the LUE-75 linear accelerator at AANL. 
The transmittance of the crystals in the wavelength range of 200-1000 nm was measured before and after irradiation, and after thermal annealing. 
The irradiation was performed in two stages. First, both crystals were irradiated with beam current of 0.125 $\mu$A, each for a total exposure time of 720 s and absorbing $5.6 \times 10^{14}~e^-$. 
Strong degradation of the optical properties of TF-1 caused by this amount of radiation was observed, while the effect on PbWO$_4$ was negligible (it is a few tens of times radiation harder than TF-1). 
In the second stage, only the PbWO$_4$ crystal was exposed to radiation, with a beam current of 0.28 $\mu$A and an exposure time of 1200 s, absorbing an additional $2.1 \times 10^{15}e^-$, still no notable effect.
Thermal annealing was performed in the temperature range of 160 to 250$^\circ$C (isochronal) for 10-12 hours. The transmittance of the annealed crystals increased with the annealing temperature and time.

\end{abstract}



 
\maketitle
\flushbottom

    
\section{Introduction}
\label{intro}

Calorimeters are key detectors for particle identification and energy measurements in high-energy physics experiments. Electromagnetic Calorimeters (EmCals) can provide good energy resolution and high-level particle identification capability for showering particles (such as $e$, $\gamma$, and decay products of $\pi^0$). 

Before the 1990s, the primary heavy crystals used in large-scale EmCals in high-energy physics were Sodium Iodide (NaI), Cesium Iodide (CsI), bismuth germinate (BGO), and Lead Glass (PbO). 
Calorimeters of lead glass types SF5, F8, TF-1, and TF-101 have been successfully used in previous high-energy physics experiments at Fermilab~\cite{FLAB-E705}, CERN~\cite{GAMS}, JLab ~\cite{HM-NIM,Puckett1,Puckett2,Puckett3}, and elsewhere.
They are still considered attractive and cost-effective alternatives to radiators in crystal-based calorimeters for future applications, with ongoing research into new scintillating glass formulations designed for enhanced radiation resistance. 

After 2000, prominent heavy crystals used in electromagnetic calorimeters for major high-energy physics experiments are Lead Tungstate (PbWO$_4$, or PWO), and newer materials, such as Lutetium 
Oxyorthosilicate (LSO/LYSO).
These materials offer high density and scintillation properties, which are necessary for precise energy and position measurements of electrons and photons. 

Radiation significantly affects all crystals in the electromagnetic calorimeters. 
Both lead glass (TF-1) and lead tungstate (PbWO$_4$) suffer radiation damage, primarily induced by the formation of color centers that degrade their optical properties. 
Damage is generally recoverable in these cases. Recovery can occur spontaneously at room temperature (known as self-recovery or room-temperature annealing), although it can be accelerated and enhanced by post-irradiation treatments, specifically thermal annealing and optical bleaching.

Lead glass exhibits moderate to good radiation hardness for calorimetric applications. However, its performance is highly dependent on the specific glass composition, impurity level, accumulated dose, and environmental conditions (such as temperature).  
The most effective method for improving the radiation hardness of lead glass is doping it with cerium ($Ce^{3+}$ ions).
Cerium-doped glasses (e.g., TF-101) can be several tens of times more radiation-hard than undoped glasses (e.g., TF-1).
The tolerable accumulated dose for undoped lead glasses is approximately $10^2$–$10^4$ rad. Cerium-doped glasses can withstand higher doses, with some of them retaining functionality up to doses $\sim2\times10^3$ rad with less than 1\% degradation.

If lead glasses offer advantages in cost and ease of production compared to lead tungstate crystals,
the latter generally exhibits superior inherent radiation hardness for specific high-intensity and high-energy physics applications, particularly when operated at low temperatures.
Lead tungstate crystals are considered the best detectors for electromagnetic showers because of their high density, small Moliere radius, fast signal, and radiation hardness. They have been used in CMS (CERN)~\cite{Marco}, PrimEx (JLab)~\cite{Kubantsev}, NPS (JLab)~\cite{NPS}, PANDA (GSI)~\cite{Erni} and elsewhere.

Lead tungstate crystals meets the basic requirements for the
electromagnetic calorimeters (EmCals) of Electron-Ion Collider 
(EIC)~\cite{EIC-Yellow}. 
EmCals are designed to withstand a "modest" radiation environment compared to the LHC, with an emphasis on using radiation-hard technologies and employing mitigation strategies such as annealing. 
The expected radiation dose for EmCals varies significantly according to the detector region (Backward, Barrel, and Forward). The levels generally range from tens of Gy to a few kGy per year for the total ionizing dose and neutron fluences from 10$^8$ to 10$^{12}$ 1-MeV neutron equivalent per cm$^2$ per year.
The highest radiation dose levels are expected in the forward direction (along the hadron beam) near the beam pipe: approximately 25 Gy (2.5 krad) per year in the insert region and up to 3 krad/year in other forward regions.

This article presents the results of our studies on the radiation hardness of two types of crystals: lead glass of the TF-1 type (produced by Lytcarino, Russia~\cite{Lytcarino}) and lead tungstate PbWO$_4$ (produced  by Crytur, Czechia~\cite{CRYTUR}). A 20 MeV electron beam from the LUE-75 linear accelerator at A. I. Alikhanyan National Science Laboratory (AANL) was used. The optical transmittance of the crystals in the wavelength range of 200-1000 nm was measured before and after irradiation and thermal annealing. 
 
The remainder of this paper is organized as follows: Section~\ref{irrad} describes the irradiation of crystals under the electron beam of the LUE-75.
In Section~\ref{damage}, details of the experimental measurements of optical transmittance and simulations quantifying radiation damage are presented. The thermal annealing process and its results are presented in Section~\ref{anneal}. 
 Finally, Section~\ref{summary} presents a summary and an outlook.


\section{Irradiation of TF-1 and PbWO$_4$ crystals under 20 MeV electron beam}
\label{irrad}

Radiation, particularly high-energy particles such as gamma rays, neutrons, and charged hadrons, introduces defects in the crystal lattice structure, primarily in the form of color centers (radiation-induced absorption bands). 
These defects create new light absorption paths, which significantly reduce the crystal transparency and, consequently, the light attenuation length and light output.
This reduction in light output impairs the ability of the detector to accurately measure the particle energies, thereby degrading the energy resolution of the calorimeter.
   
Before irradiation, the crystals were visually inspected, and their transverse transparency was measured using a 402 OCEAN-ART (FLAME-S-XR1) optical spectrometer. Its operational range encompasses ultraviolet to visible light
(UV-Visible, 200-1025 nm). We used the OceanArt Software application for data analysis, which allows for the measurement of absorption, transmittance, and reflection.
The measurement process includes the following three main steps:

- Initial measurement (performed with the light on without the crystal).

- Dark current measurement (with the light on, but the shutter is closed).

- Transparency measurement (with the shutter open, crystal in place, between the input and output fibers).

For each crystal, measurements were taken at 12 equidistant points along the crystal, separated by 1 cm from each other, starting from 1 cm deep from the front-radiated surface. Each measurement was repeated 3-5 times, and the obtained data were averaged.
By examining the variations in the results, we estimated an error of $\sim$2\% ($\sim$10\%) for wavelengths above (below) 380 nm (see Section~\ref{transmit} for more details).

The crystals were then moved into the accelerator's experimental hall and installed in front of the $\sim$14 mm diameter collimator exit of the analyzing magnet (MA) at a distance of approximately 15–20 mm from the exit point.
They were mounted on a remote-controlled table that could be used to remove or place them in front of the beam.
Because the TF-1 and PbWO$_4$ blocks were of different sizes, they were vertically adjusted in front of the beam so that in both cases, the beam fell into the center of the front face of the crystal.  
The crystals were irradiated at an ambient temperature of $\sim20^\circ$C. 
After irradiation, the crystals were stored at the accelerator site for approximately 80 h to reduce residual activity.
More details on the accelerator and irradiation are provided in the following sections.

\subsection{Linear accelerator LUE-75}
\label{LUE}

The LUE-75 linear electron accelerator of the Experimental Physics Division of the AANL is the only operating electron accelerator in Armenia in the energy range of up to 75 MeV. Currently, there are no charged-particle accelerators with energies in this range in neighboring countries. 

The LUE-75 resonant electron accelerator on a ten-centimeter traveling wave served as the injector of the Yerevan 6 GeV synchrotron ARUS. The operation of the annular part of the ARUS is currently suspended, and the LUE-75 functions autonomously as the basic complex to study urgent problems in low-energy nuclear physics in the 10–75 MeV energy range. 

A thermionic gun with Pierce optics was used as the source of electrons in the LUE-75, from the output of which a 50 keV beam entered the input of the waveguide buncher (in the injection section). 
In the buncher, during autophasing, the electron flux, which is uniformly distributed over the wave phases, is grouped into bunches (modulated in density) with a repetition rate equal to the frequency of the microwave accelerating field and is simultaneously accelerated to an energy of 3 MeV. Further acceleration is performed by three identical main accelerating sections that receive microwave energy from three powerful klystron posts. 

\begin{figure}
\centering
\includegraphics[width=0.45\linewidth,height=5.0cm, angle=0.0]{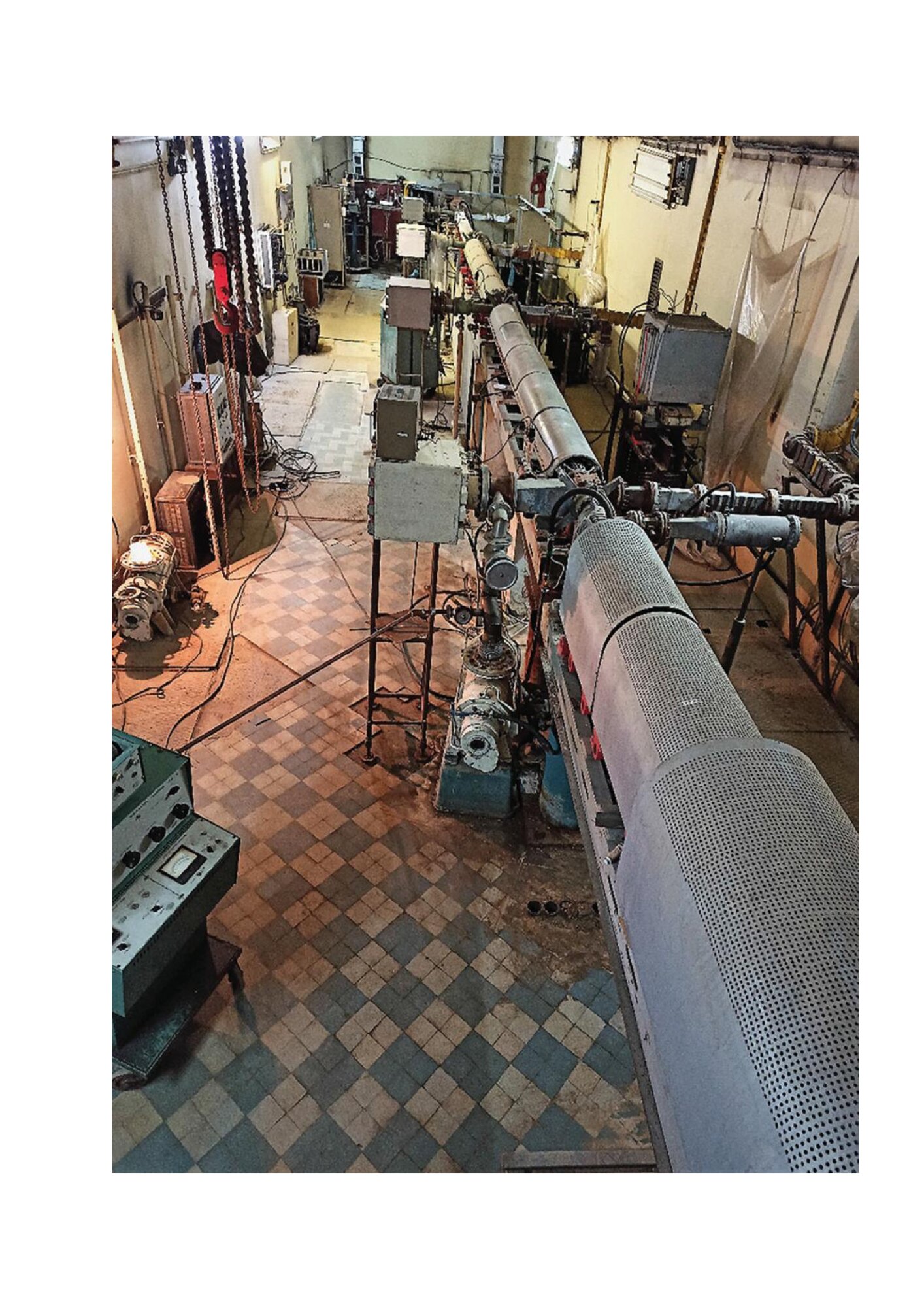} 
\includegraphics[width=0.45\linewidth, height=5.0cm,angle=0.0]{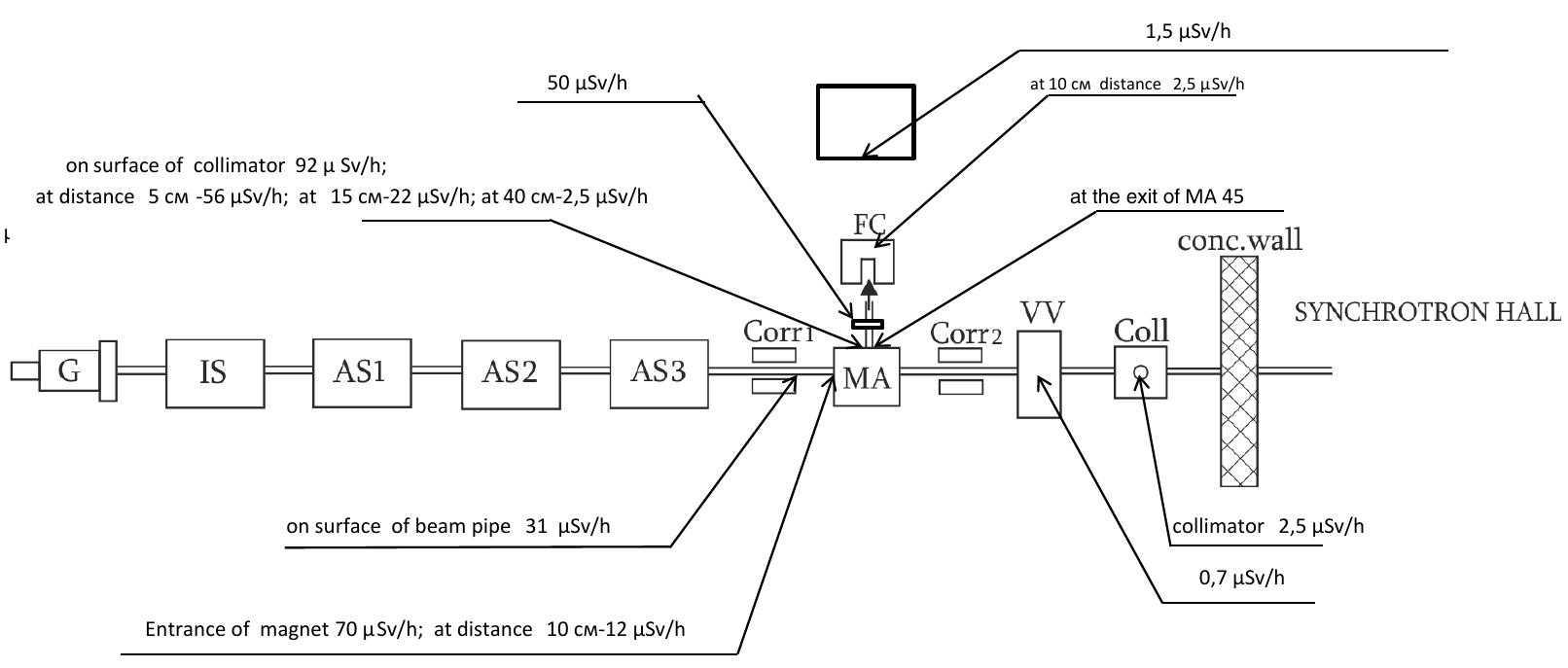}
\caption{\label{fig:LUE-75} (Left) The accelerating sections of the LUE-75, and (Right) schematic diagram of the residual radiation map in the test area after 2 weeks shutdown. }
\end{figure}

The first klystron operates according to the self-excited oscillator scheme proposed by AANL specialists in the 1970s~\cite{LUE-75}, supplying the remaining klystrons operating in amplifier mode with input power. The advantages of the LUE-75 self-excited oscillator power supply system are its operational reliability, simplicity, and lack of an expensive external driver. It is possible to operate in the mode of external excitation of powerful amplifying klystrons if necessary.
The nominal trigger pulse repetition rate was set to 50 Hz. Figure~\ref{fig:LUE-75} shows the accelerating sections of LUE-75 and a schematic diagram of the residual radiation map in the test area after two weeks of shutdown.
The accelerator provides experimenters with electron beams with an average current of up to 10 $\mu$A (without collimation), which corresponds to a pulse current of up to 150–200 mA, depending on the macropulse duration and energy.
The electron beam direction is almost perpendicular to the Earth’s magnetic field lines, and extended correction coils are installed to neutralize the influence of the geomagnetic field on the electron beam.

The LUE-75 linear accelerator complex includes the linear accelerator itself and the parallel transport path located in the annular hall of the synchrotron, away from the LUE-75 hall, and behind its radiation shielding wall. A significant reduction in both the radiation background when the synchrotron was turned off and the influence of various electromagnetic interferences and noise from the linear accelerator's electrical and radio-technical devices on the measuring equipment was achieved. The LUE-75 accelerator is specifically noted for conducting precision experiments, particularly in irradiation and low-energy nuclear physics applications. 

The beam parameters and their stability, which determine the quality of the experiment, depend on several factors. With small deviations of the initial phase from the optimal one, the energy spread of the beam is approximately the same for all three sections with different electron energies at the inputs; thus, a phase change by $\pm8^\circ$ from the optimal value results in a relative change in energy of $\pm$1\%. 
A change in temperature by $\pm 0.5^\circ$C results in a change in the initial phase $\pm1^\circ$. Such a deviation of the initial phase from the optimal one introduces an insignificant change in energy, approximately 0.014\%.

The additional broadening of the energy spectrum also occurs because of the dispersion properties of the waveguide. 
A collimated and well-bunched beam at nominal energies has an energy spectrum width (FWHM) of approximately 2\%.


\subsection{Irradiation of crystals}
\label{ss_irrad}

\begin{figure}
\centering
\includegraphics[width=0.45\linewidth,height=5.0cm, angle=0.0]{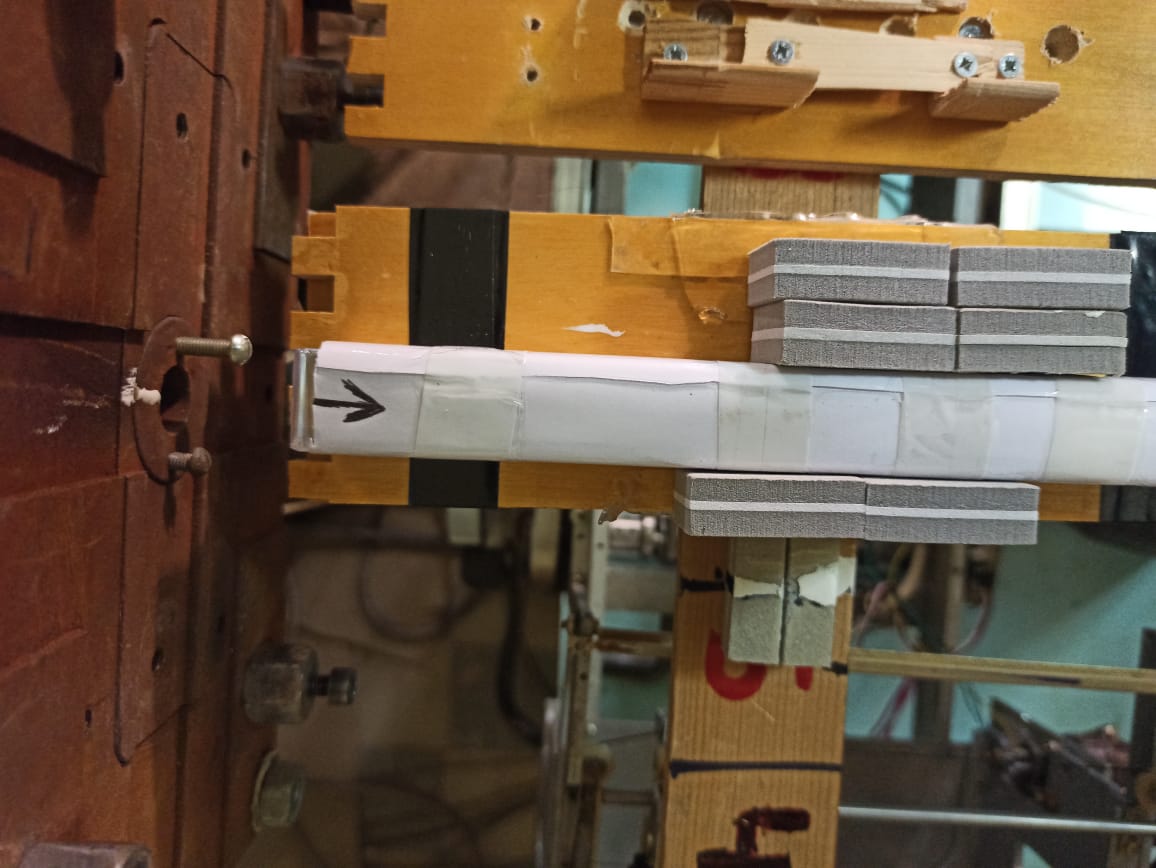} 
\caption{\label{fig:pbwo-on-beam} The PbWO$_4$ crystal installed at the front of collimator of the LUE-75.}
\end{figure}

Irradiation was performed in two stages. First, both crystals were irradiated with an electron beam current of 0.125 $\mu$A, for a total exposure of 720 s, absorbing 
$N_e\approx5.6\times10^{14}~e^-$.
In the second stage, only the PbWO$_4$ crystal was exposed to additional radiation, the beam current was 0.28 $\mu$A, and the exposure time was 1200 s, resulting in the absorption of additional $N_e\approx2.1\times 10^{15}~e^-$.
Figure~\ref{fig:pbwo-on-beam} shows the PbWO$_4$ crystal installed in front of the collimator for irradiation. 

A Faraday cylinder was located approximately 50 cm from the exit collimator of the analyzing magnet, which was used to periodically monitor the beam current as the crystal was moved out.
Beam profile measurements indicated that in both cases more than 85-90\% of the electrons were absorbed in the crystals. Hence, we assigned an uncertainty on the number of incident electrons in our subsequent estimates of the radiation dose at the level of 10-15\%.

The chemical compositions of the crystals are significantly different.
TF-1 consists of PbO (51.2\%), SiO$_2$(41.3\%), K$_2$O(3.5\%), Na$_2$O(3.5\%) (see Ref.\cite{Lytcarino,HM-NIM}); while  Crytur PbWO$_4$ are typically pure (99.995\%) scintillator materials, mixture of PbO (50\%) and WO$_3$(50\%) (see Ref.\cite{CRYTUR,Burachas}). 
Their dimensions, densities, radiation lengths, and Moliere radii are 40$\times$40$\times$400~mm$^3$, $\rho\approx$3.86 $g/cm^3$, $X_0\approx$2.7 cm, and $R_M\approx$4.0 cm  for TF-1, and 20$\times$20$\times$200~mm$^3$, $\rho\approx$8.3 $g/cm^3$, $X_0\approx$0.9 cm, and $R_M\approx$2.2 cm for PbWO$_4$, respectively.

We decided to start irradiation with the TF-1 block.
To avoid thermal damage to the crystal, a large accumulated dose was achieved by radiation in short periods.
In the first stage, the beam current was 0.125 $\mu$A, and the total exposure time for both crystals was $\tau$=720 s. The exposition was performed in six periods of 120 s each, followed by a 120 s pause. The accumulated total energy was ${\triangle E = N_e\times E_e\times\tau\approx}$ 1800 J.

Strong degradation of the optical properties of TF-1 caused by this amount of radiation was observed, while the effect on PbWO$_4$ was negligible (it is a few tens of times radiation harder than TF-1). Therefore, we performed additional exposition for PbWO$_4$.

In the second stage, only the PbWO$_4$ crystal was exposed to radiation, with a beam current of 0.28 $\mu$A and an exposure time of 1200 s, absorbing an additional energy $\triangle E\approx4.2\times 10^{16}$~MeV$\approx$6700 J, still no notable effect.

\begin{figure}
\centering
\includegraphics[width=0.45\linewidth,height=2.5cm,angle=00.0]{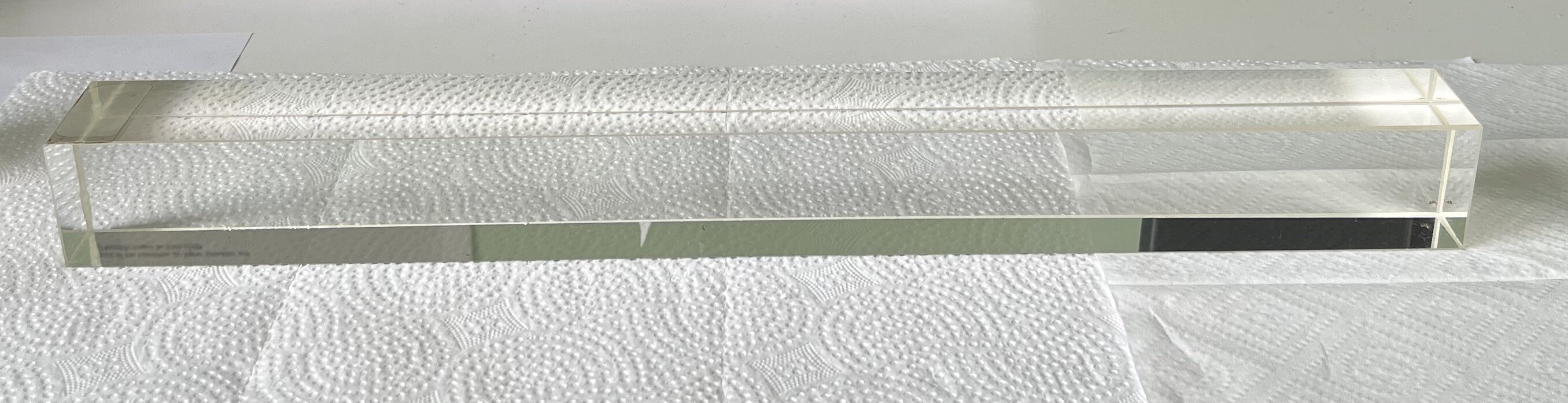}
\includegraphics[width=0.45\linewidth,height=2.5cm,angle=0.0]{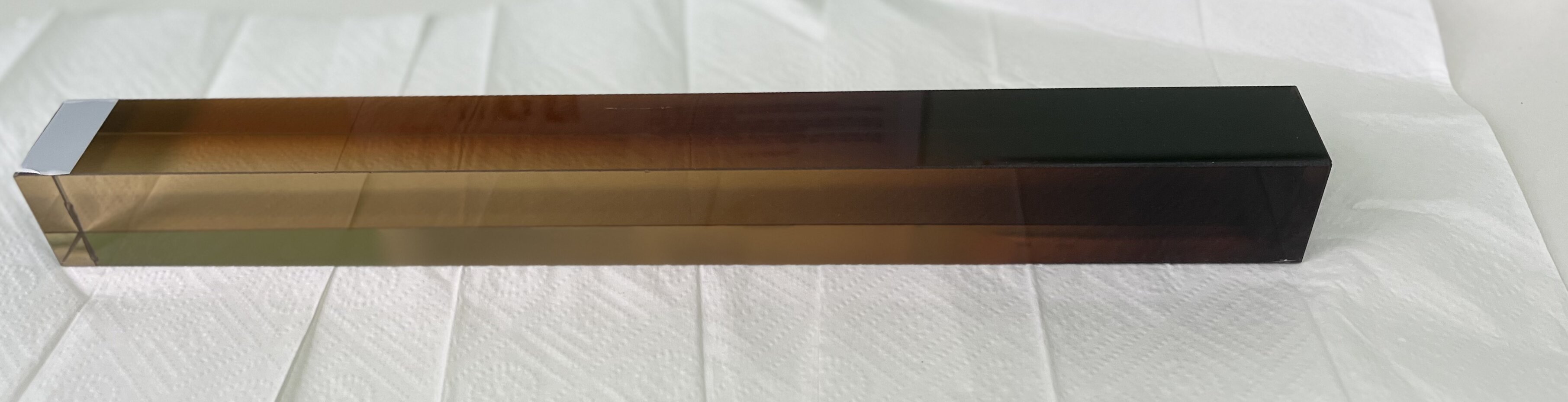}
\includegraphics[width=0.45\linewidth,height=2.5cm,angle=0.0]{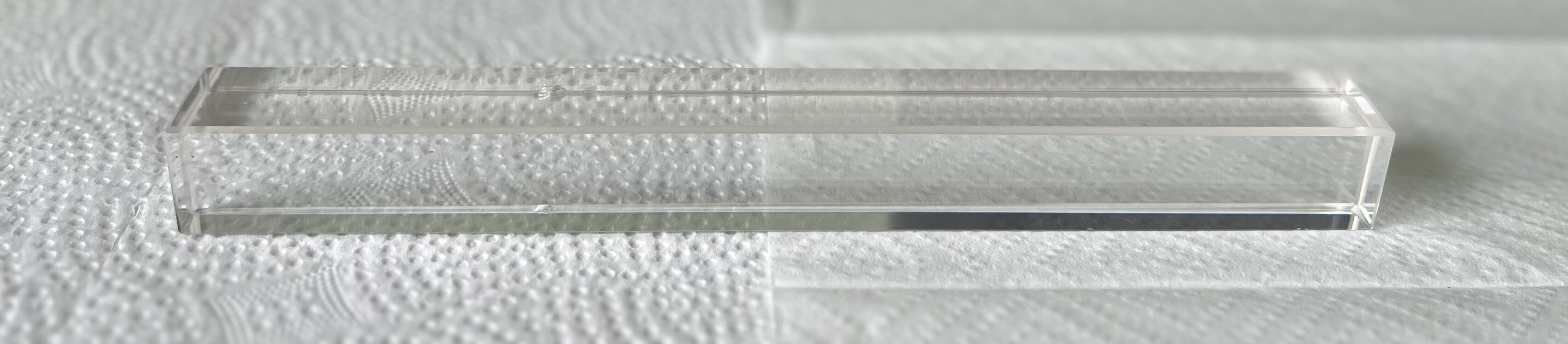}
\includegraphics[width=0.45\linewidth,height=2.5cm,angle=0.0]{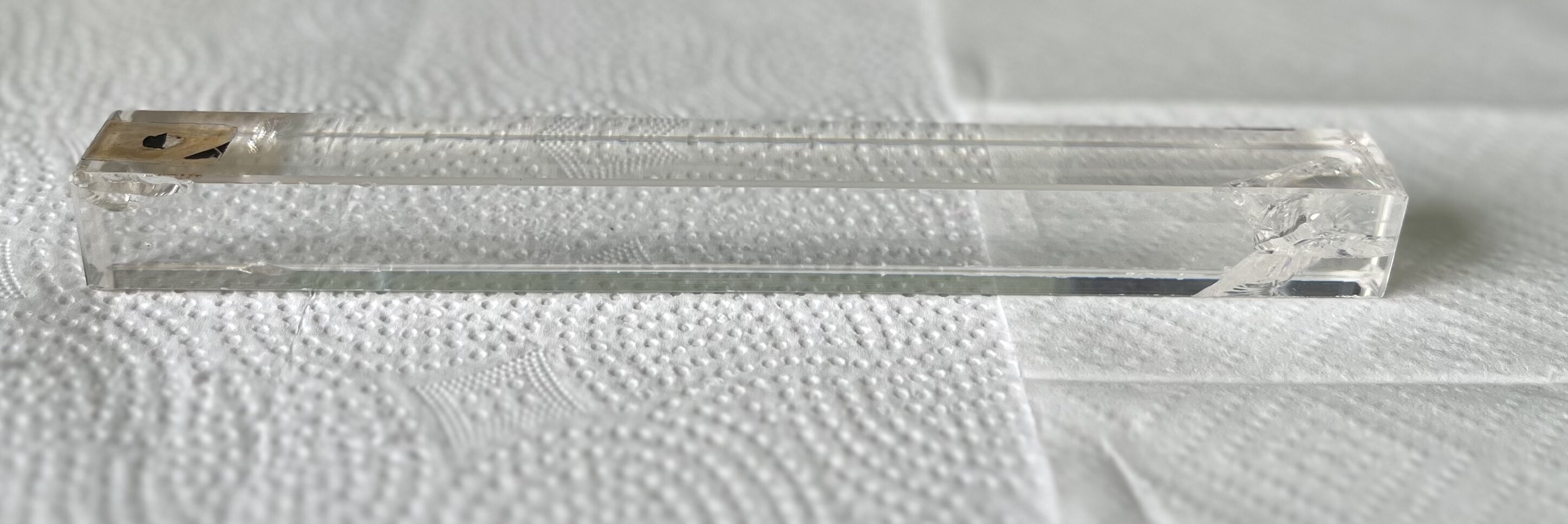} 
\caption{
Lead glass (TF-1) and lead tungstate (PbWO$_4$) crystals before (top and bottom left) and after (top and bottom right) irradiation with electrons of energy 20 MeV. The number of electrons absorbed by the crystals was  $\sim5.6\times10^{14}~e^-$ for TF-1 and $\sim2.66\times10^{15}~e^-$ for PbWO$_4$ (accumulated after two steps of irradiation). }
    \label{fig:TF1-PbWO-irrad}
    \end{figure}

 Figure ~\ref{fig:TF1-PbWO-irrad} shows the  lead glass and lead tungstate crystals before (top left and bottom left) and after (top right and bottom right) irradiation. The number of electrons absorbed by the crystals was  $\sim5.6\times10^{14}~e^-$ for TF-1 and $\sim2.66\times10^{15}~e^-$ for PbWO$_4$ (accumulated after two stages of irradiation).
 One can see that the effect of irradiation on TF-1 is essential, specifically on the front 3-4 cm of the crystal, whereas it is negligible on the PbWO$_4$ crystal.

Crytur's lead tungstate crystals for calorimetry are doped, most commonly with lutetium (Lu), lanthanum (La), yttrium (Y), or niobium (Nb). Doping is an essential technology for improving several critical properties of crystals for use in high-energy physics experiments, such as those at the CERN Large Hadron Collider (LHC) or PANDA at GSI.
This may explain why we did not observe a color change after irradiation with PbWO$_4$.

During the second stage of irradiation, after the first 120 s, the front $\sim$3 cm section of the PbWO$_4$ crystal cracked owing to a nonuniform thermal expansion. However, we continued the planned irradiation to observe the degradation of the transverse transmittance beyond 3 cm from the front. 
 

\section{Quantifying the radiation Damage}
\label{damage}

Before the crystal was dismounted from the support stand and removed from the accelerator hall, its residual activity was measured. For both crystals, we found the residual activity approximately 2-3 times higher than the natural radioactivity (below 5-6 mSv).  
 
The fraction of energy absorbed from 20 MeV electrons in 3.0 radiation lengths is approximately 0.95 (or 95\%), which is $\approx$8 cm for TF-1 and $\approx$3 cm for PbWO$_4$.
The beam spot on the face of the crystal was radially symmetric with a Gaussian distribution
and a standard deviation $\sigma\sim$3.7-3.8 mm, conditioned by a collimator with a diameter of $\sim$14 mm. 
If we assume that the energy of the electrons was absorbed by the material within the volume with a surface area of the crystal and a depth of three radiation lengths, then the corresponding masses for TF-1 and PbWO$_4$ would be $M_{TF1}\approx$0.495~kg and $M_{PWO}\approx$0.100~kg.
The integral radiation-absorbed dose for the first stage of irradiation of TF-1 and PbWO$_4$ would be D$\approx$ 0.36 Mrad and 1.8 Mrad\footnotetext[1]{1 rad = 0.01 Gy = 0.01 J/kg, or 1 J/kg = 100 rad}, respectively.
In the second stage of irradiation, PbWO$_4$ accumulated an additional dose D$\approx$6.7 Mrad\footnotetext[2]{D=$\triangle E/m$}.

However, our MC simulations show that for the tested crystal widths, and within the first 3 radiation lengths of the EM shower development, the energy leakage occurs at about $\sim$25\% for TF-1 and $\sim$31\% for the PbWO$_4$ (see Subsection ~\ref{MC_sim}).
Therefore, considering this, the actual radiation dose received by the TF-1 and PbWO$_4$ crystals for the first stage of irradiation will be $\sim$0.27 Mrad and $\sim$1.2 Mrad, respectively, and during the second irradiation, the PbWO$_4$ received an additional $\sim$4.6 Mrad, or a total of $\sim$5.8 Mrad.

The specific heat capacity of lead tungstate at room temperature is approximately 0.263 J/g/$^\circ$C. Therefore, the deposited power will increase the temperature by about 0.20-0.25$^\circ$C/s, or approximately 20-25$^\circ$C within 2 min of irradiation, if we assume uniform irradiation on the first 3 cm frontal section of the crystal. However, because of the non-uniformity of irradiation, the temperature increase in some areas of the crystal was probably much higher, which caused a crack in the crystal (see the bottom panel of Figure~\ref{fig:TF1-PbWO-irrad}).
We observed a similar effect in 2016 during the irradiation of crystals in the Idaho electron accelerator when the dose intensity exceeded 1.3 Mrad/h (see Ref.~\cite{Horn}).

The results of our study indicate a strong degradation of the optical properties of TF-1 after absorption of $5.6 \times 10^{14}~e^-$,
and no notable color changes in PbWO$_4$ 
after the absorption of $2.66 \times 10^{15}~e^-$, confirming that PbWO$_4$ is more resistant to radiation than TF-1 does.  
 

\subsection{Transmittance}
\label{transmit}

\begin{figure}
\centering
\includegraphics[width=0.45\linewidth,height=5.0cm,angle=0.0]{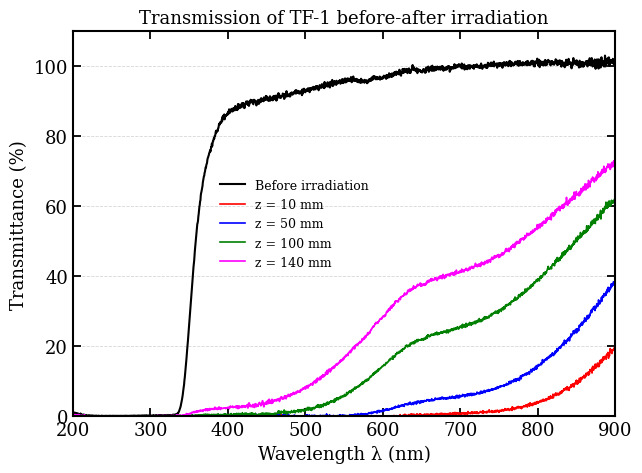}
\includegraphics[width=0.45\linewidth,height=5.0cm,angle=0.0]{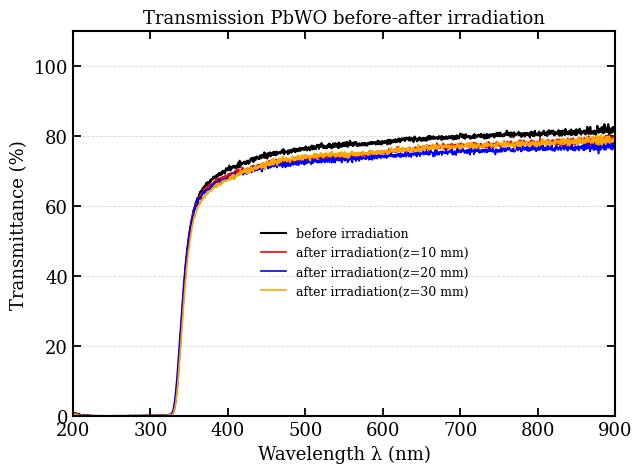} 
    \caption{ Transmittance of Lead glass (TF-1) and lead tungstate (PbWO$_4$) blocks as a function of wavelength $\lambda$.The black curves represent the data before irradiation. The colored curves show the transmittance after irradiation with electrons ($\sim$5.6$\times10^{14}$ for TF-1 and 
    $\sim$2.66$\times10^{15}$ for PbWO$_4$) at different distances from the front of the crystals (10, 50, 100, and 140 mm for TF-1 and 10, 20, and 30 mm for PbWO$_4$). No corrections were made for the reflections from the crystal surfaces.  }  
    \label{fig:TF1-PbWO-trans}
    \end{figure}

Figure~\ref{fig:TF1-PbWO-trans} shows the transmittance of the lead glass and lead tungstate blocks as a function of the wavelength $\lambda$.
The black curves represent the data before irradiation. The colored curves show the transmittance after irradiation with electrons of energy 20 MeV 
 ($\sim$5.6$\times10^{14}~e^-$ for TF-1 and  $\sim$2.66$\times10^{15}~e^-$ for PbWO$_4$).
Our measurements conform to the conclusion of Ref.\cite{Schaefer}, when the glass (F8 crystal) is irradiated, the transmittance (and attenuation length) in the short wavelength region of the spectrum is degradated more than in the red portion of the spectrum\footnotetext[3]{ In Ref.\cite{Schaefer} the damage in the attenuation length of F8 (composition similar to TF-1) irradiated with 20 MeV energy electrons was measured.}.

The measurements after irradiation were performed at different distances from the front of the crystals (10, 50, 100, and 140 mm for TF-1 and 10, 20, and 30 mm for PbWO$_4$).
No corrections were made for the reflections on the crystal surfaces.

It should be noted that our results for the transmittance of TF-1 and PbWO$_4$ before irradiation are quite close to those reported in other studies (for example, \cite{HM-NIM}, \cite{NPS} and \cite{HM-EPJ,HM-Cont}).
 A strong coordinate dependence of TF-1 was evident from the measurements taken after irradiation.
 This is because of the different amounts of energy deposited by electrons with depth in the crystal, which can be estimated using simulations.
 Owing to the high radiation resistance of the PbWO$_4$ crystal, this effect was insignificant at this dose.


\subsection{Monte Carlo Simulation}
\label{MC_sim}

Monte Carlo simulations were performed using the Geant4 package~\cite{Geant4}, version 11.2.2. A Geant4 reference Physics List FTFP\_BERT, was utilized,  which included all the required parameters for the accurate modeling of the development of electron-induced showers in the crystals in electromagnetic processes, such as bremsstrahlung and pair production. 
 
The accuracy of the simulations depends on the adequate description of the chemical composition and densities of the crystal materials. The PbWO$_4$ crystal was modeled according to its chemical formula; its density was 8.3 g/cm$^3$ (see subsection \ref{ss_irrad}).  The small amounts of doping present in the new generation of PbWO$_4$ crystals (known as PWO-II-type) were not considered in this study. The chemical composition of TF-1, with a density of 3.86 g/cm$^3$, was also considered, as presented in subsections \ref{ss_irrad} and \cite{HM-NIM}.

The beam conditions were modeled based on the LUE-75 studies (see Subsection \ref{LUE}) and our rough measurements of the beam cross-section at the crystal position before the irradiation. The beam energy was sampled from the normal distribution of the nominal 2\% FWHM spread. The beam profile was radially symmetric with a Gaussian distribution and a standard deviation equal to one-quarter of the diameter of the collimator. The angular divergence of the beam after collimation was ignored.

\begin{figure}
\centering
\includegraphics[width=0.45\linewidth,height=5.0cm, angle=0.0]{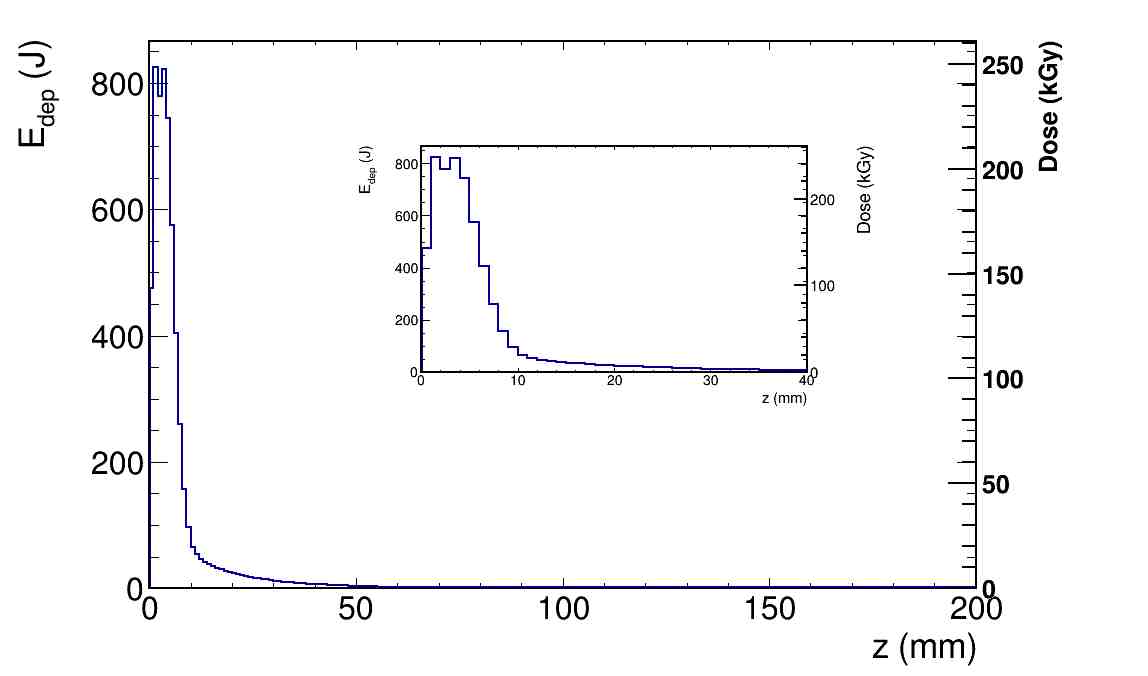} 
\includegraphics[width=0.45\linewidth,height=5.0cm, angle=0.0]{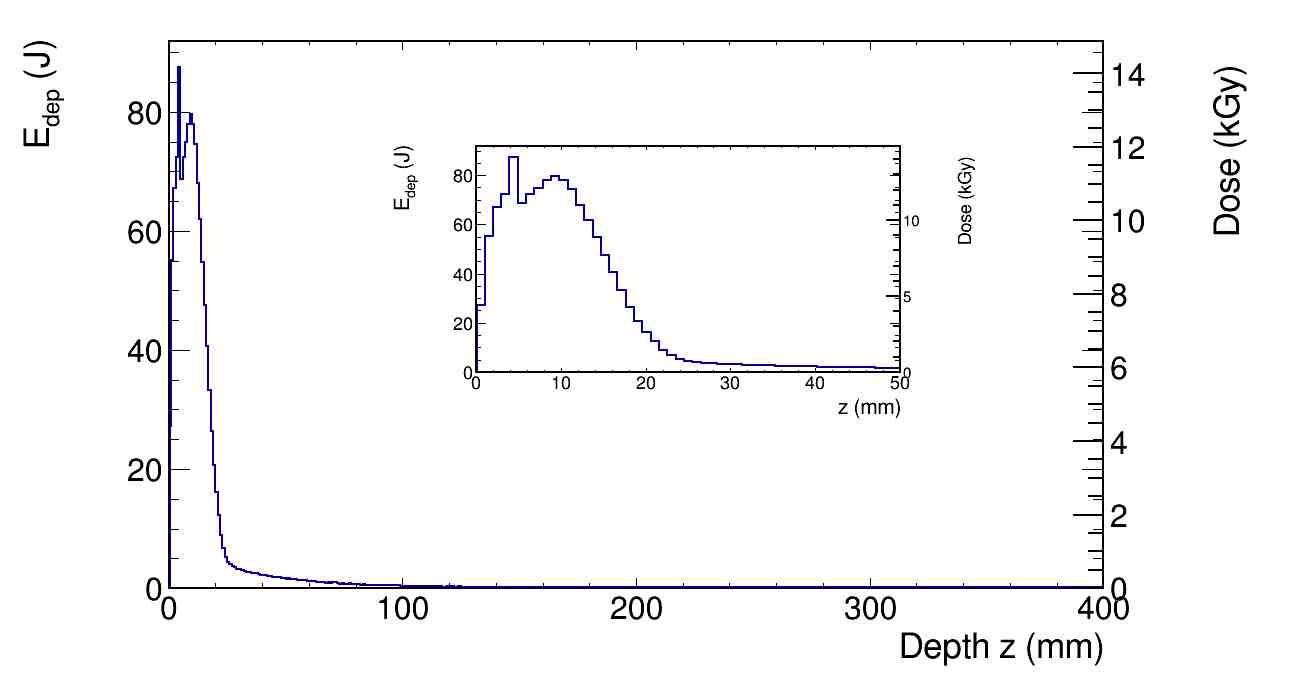}
\caption{\label{fig:Edep_long} Longitudinal energy deposition profiles in the 20 cm long PbWO$_4$ crystal (left) and 40 cm long TF-1 type lead glass block (right) from the GEANT4 simulations. See text for details.}
\label{fig:TF1-PbWO-edep}
\end{figure} 

The longitudinal profiles of the accumulated energies in the crystals obtained from the simulations are shown in Figure~\ref{fig:TF1-PbWO-edep}. The distributions were scaled to the total number of electrons delivered from the beam during irradiation ($2.66\times 10^{15}~e^-$ for PbWO$_4$ and $5.62 \times 10^{14}~e^-$ for TF-1). As shown, the main energy fraction was deposited in the first radiation length of the crystals, followed by a gradual decrease. Note that, in the energy range above 1 GeV, the position of the peak deposition depends logarithmically on the energy of the primary particle \cite{PDG}.

As shown (see Figure~\ref{fig:TF1-PbWO-edep}), for both cases, the media were deep enough (22.4  and 14.6 radiation lengths for PbWO$_4$ and TF-1, respectively) to exclude tangible energy leakage from the back of the crystal at these low energies. In contrast, the crystals are only approximately one Moliere radius thick transversely, and there is a significant beam spread in the transverse plane. 
Hence, the energy leakage of $\sim$29\% for the PbWO$_4$ crystal and {$\sim$20\%} for the TF-1 block was mostly in the transverse direction.
Note that for the lesser region of the bulk energy absorption, within approximately the first three radiation lengths (3 cm in PbWO$_4$ and 8 cm in TF-1), the estimates of energy leakage are larger, $\sim$31\% for PbWO$_4$ and $\sim$25\% for TF-1.

The estimates for the peak doses of radiation from the simulations were $\sim$250 and $\sim$12 kGy for the PWO crystal and TF-1 block, respectively. The difference of an order of magnitude correlates with the accumulated beam intensity during the irradiation.
Our MC calculations qualitatively reflect the deterioration of the optical transmittance of the crystals (especially for Tf-1) under irradiation (see Figure~\ref{fig:TF1-PbWO-trans}).
  

\section{Thermal Annealing} 
\label{anneal}

Some crystals, such as BGO and PbWO$_4$, show a degree of spontaneous recovery from gamma radiation damage at room temperature over a period of hours to weeks, making the damage dose-rate dependent.
In oxide crystals (e.g., TF-1, PbWO$_4$, and BGO), damage is often caused by defects in the intrinsic structure, such as oxygen vacancies in the crystal lattice.

Methods for the recovery of radiation-damaged crystals include thermal annealing (heating to high temperatures), optical bleaching/stimulation (using specific light wavelengths, such as blue/IR, to depopulate color centers), and complex combined laser/thermal treatments. Recovery varies significantly by crystal type (e.g., TF-1, PbWO$_4$, and BGO) and damage mechanism, sometimes requiring in situ monitoring and offering only partial restoration. 

Thermal annealing is a heat treatment process that alters the physical and sometimes chemical properties of a material by heating it to a specific temperature, holding it at that temperature (soaking), and cooling it at a controlled rate. 

Low-temperature thermal annealing of crystals is primarily a recovery process aimed at removing defects and relieving internal stress without causing significant grain growth or massive restructuring (recrystallization) of the crystals. At these lower temperatures, atoms possess sufficient kinetic energy to migrate slightly within the lattice, a process driven by the desire to reach a lower energy, more stable equilibrium state. 

High-temperature thermal annealing is a heat treatment process that allows atoms within a crystal to gain mobility, enabling the material to transition from a high-energy defective state to a lower-energy equilibrium crystalline state. The primary processes that occur are recovery, recrystallization, and grain growth, which collectively reduce lattice defects (such as dislocations and vacancies) and increase ductility. 

During this process, the crystal structure undergoes significant changes to reduce defects, relieve internal stress, and improve ductility. 
Thermal annealing of crystals, typically conducted in air at high temperatures, acts as a purification and repair process to modify intrinsic defects, improve optical transmittance, and enhance radiation hardness. During this process, the crystal undergoes structural relaxation, oxygen exchange with the atmosphere, and reorganization of color centers.
The annealing mechanism in PbWO$_4$ is primarily the diffusion of oxygen and modification of oxygen-related defects, which remove the absorption band at a wavelength of 350 nm and significantly enhance the crystal resistance to radiation damage~\cite{Zhu-2002}.

The annealing treatments of a PbWO$_4$ crystal sample were carried out in air, sequentially from 640 to 1040°C in~\cite{Han-1999}. 
The experimental results show that annealing of PbWO$_4$ crystals at temperatures above 740°C in air can effectively wash out the intrinsic color centers, causing a 350 nm optical absorption band and improving the radiation hardness of PbWO$_4$ crystals. In addition, during the radiation procedure, the conversion of the 350 nm intrinsic color centers to the 410 nm temporary color centers was observed experimentally.
Unfortunately, these important data are for old-generation crystals (known as PWO-I), and information on new crystals (PWO-II) produced by Crytur is limited.


\subsection{Procedure of annealing}
\label{anneal-proced}

The thermal annealing procedure for lead glass (TF-1) or lead tungstate (\(\text{PbWO}_{4}\)) crystals involves specific heating and cooling cycles in a controlled atmosphere to reduce crystal defects, relieve internal stresses, and improve properties such as the light output and radiation hardness. The exact conditions (temperature, duration, atmosphere) depend on the desired outcome and initial conditions of the crystal (e.g., as-grown, doped, or irradiated). 

As an example, in Ref.~\cite{Yang}, a PbWO$_4$ crystal grown using the Czochralski method was annealed under different conditions. The transmittance of PbWO$_4$ decreased when annealed in an oxygen-rich environment, and increased when annealed in a vacuum.
The effectiveness of thermal annealing is dependent on the temperature, time, and initial radiation dose. Full recovery can take days or weeks at room temperature after significant irradiation; however, heating can expedite this process.

\begin{figure}
\centering
\includegraphics[width=0.45\linewidth,height=7.4cm,angle=0.0]{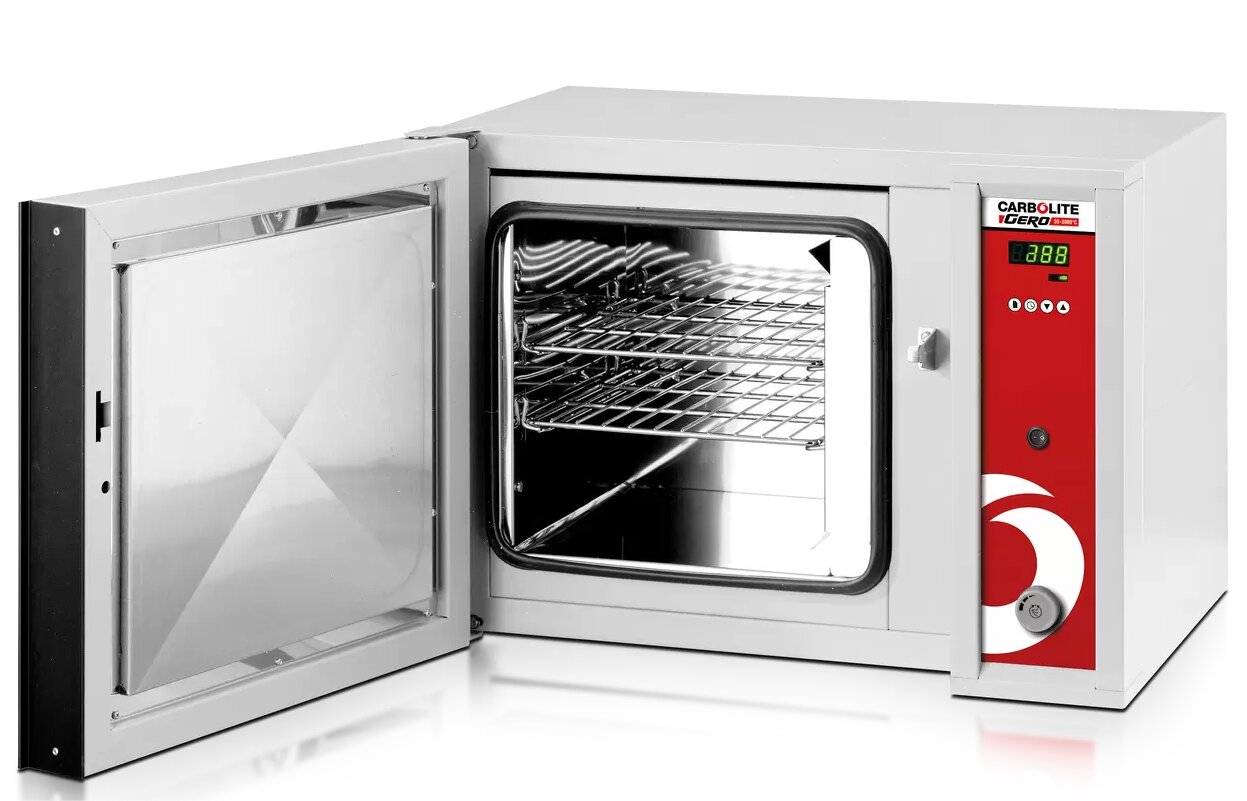}
\includegraphics[width=0.45\linewidth,height=7.4cm,angle=0.0]{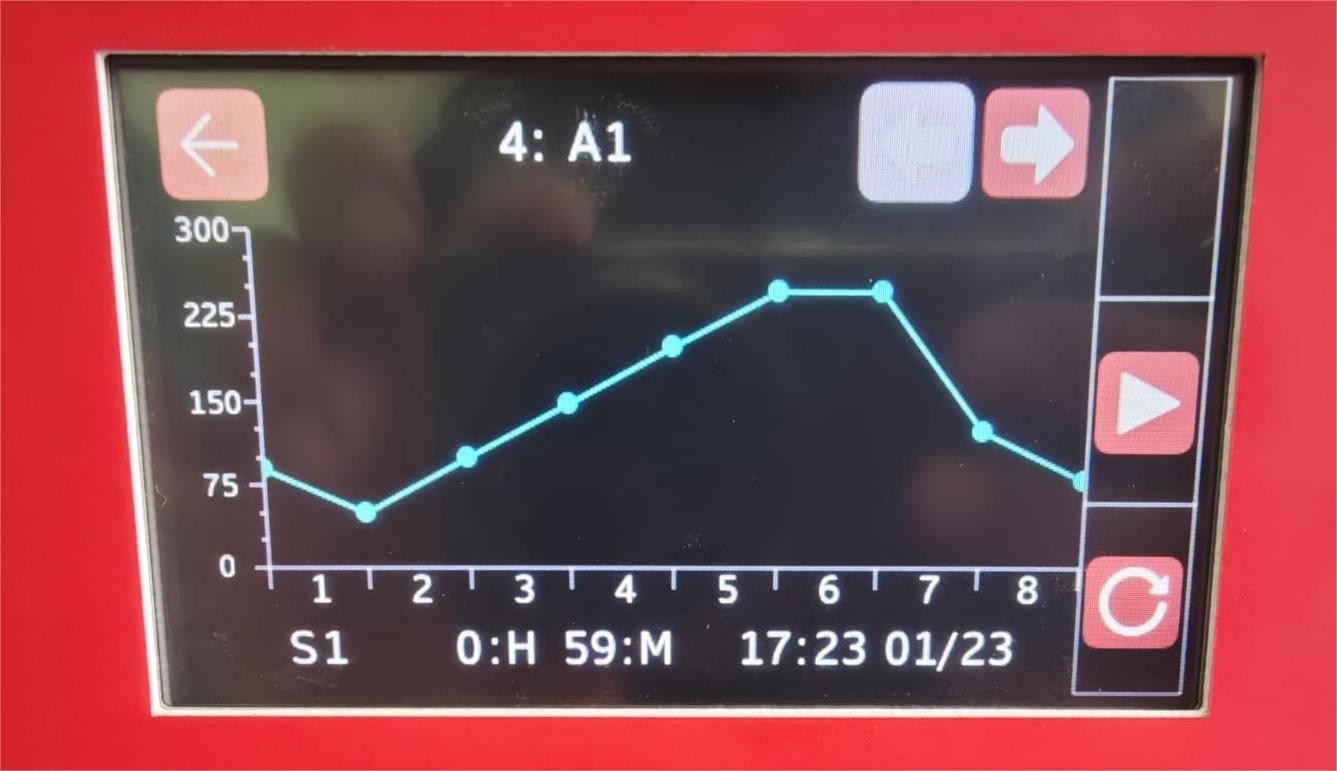}
    \caption{ The oven Carbolite and its display for one of the thermal annealing modes.} 
    \label{fig:oven-anneal}
    \end{figure}

For the thermal annealing of the crystals, we used a Carbolite Gero oven. Figure~\ref{fig:oven-anneal} shows the oven and its display for one of these heating modes.

In general, annealing involves three key steps: heating the material to a specific high temperature (often above recrystallization), soaking (holding it there) to allow microstructural changes such as stress relief and grain reorganization, and then cooling it slowly and in a controlled manner, usually in a furnace, to achieve the desired properties such as increased ductility, softness, or improved crystallinity by refining the internal structure.

  It allows for the control of a gradual increase in temperature from room temperature to a target value (maximum of 300°C) at a programmed rate, maintaining that temperature as long as selected, and gradually lowering it.

The thermal annealing of TF-1 was performed in two stages. 
In the first stage, the temperature was raised from room temperature to 160$^\circ$C over a period of 6 hours, held at that temperature for 10 hours, and then gradually lowered to room temperature over a period of 6 hours.
Consequently, the optical properties of TF-1 were only partially restored. 
In the second stage, the temperature was raised from room temperature to 250°C over 8 h, held at that temperature for 12 h, and then gradually lowered to room temperature over 8 h.
After this stage, the TF-1 block was almost completely re-established. This treatment was also applied to the PbWO$_4$ crystal.
 
Annealing is a heat treatment process that restores some of the original physical properties of a material by allowing atoms to migrate within the crystal lattice, reducing the number of defects, and relieving internal stresses. 
The elevated temperature provides the necessary energy for the mobile interstitials and vacancies to annihilate each other, thereby removing the color centers.
Annealing can restore both the optical transparency and scintillation performance of the damaged crystals to their pre-irradiation state or better.


\subsection{Quantifying thermal annealing}

\begin{figure}
\centering
\includegraphics[width=0.45\linewidth,height=2.5cm,angle=0.0]{TF1-picture-after-irrad1.jpg}
\includegraphics[width=0.45\linewidth,height=2.5cm,angle=0.0]{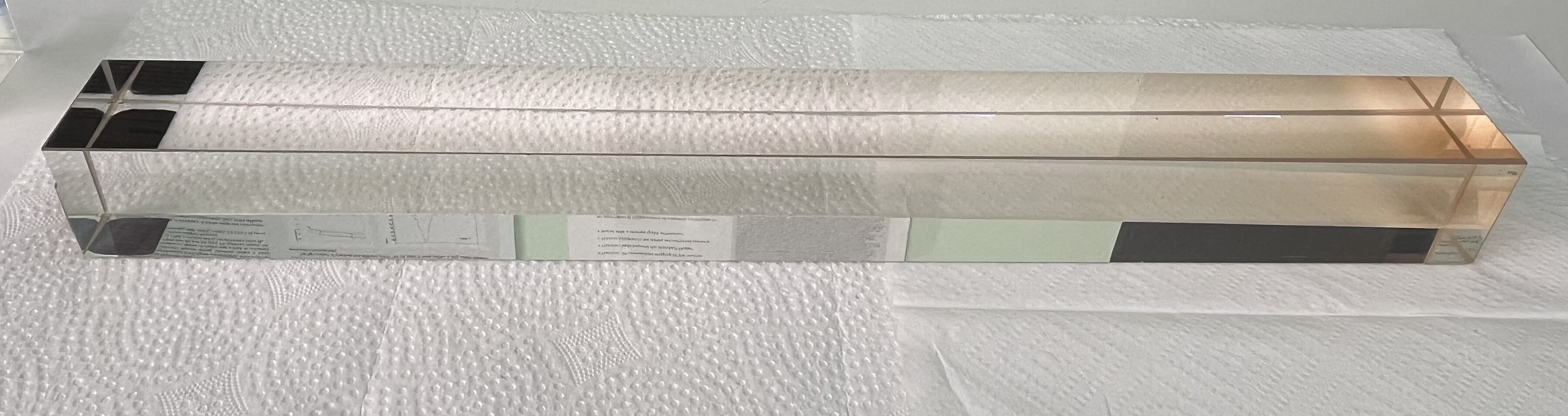}
\includegraphics[width=0.45\linewidth,height=2.5cm,angle=0.0]{pwo-picture-after-rad2.jpg}
\includegraphics[width=0.45\linewidth,height=2.5cm,angle=0.0]{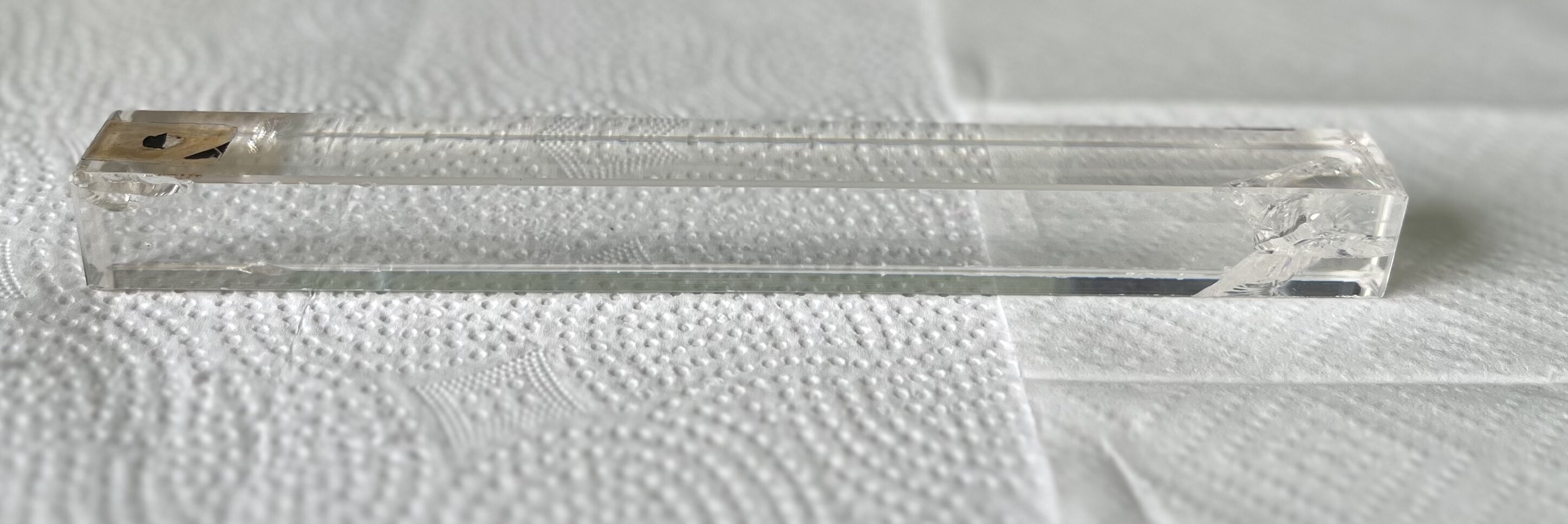} 
  \caption{ 
  (Top) Lead glass (TF-1) irradiated to $\sim$0.3 Mrad accumulated dose and after the first step of thermal annealing. (Bottom) Lead tungstate (PbWO$_4$) irradiated with $\sim$7 Mrad dose before and after thermal annealing. } 
   \label{fig:TF1-PbWO-anneal}
     \end{figure}

Figure~\ref{fig:TF1-PbWO-anneal} shows the lead glass (TF-1) block irradiated to $\sim$0.3 Mrad and the lead tungstate (PbWO$_4$) crystal irradiated to about $\sim$7 Mrad, with electrons of 20 MeV energy, before (top) and after (bottom) the first stage of thermal annealing.
The positive effect of restoring the optical characteristics in TF-1 lead glass from annealing is obvious, even visually, whereas the radiation damage and the positive effect in PbWO$_4$ from annealing are difficult to notice.

\begin{figure}
\centering
\includegraphics[width=0.35\linewidth,height=7.0cm,angle=0.0]{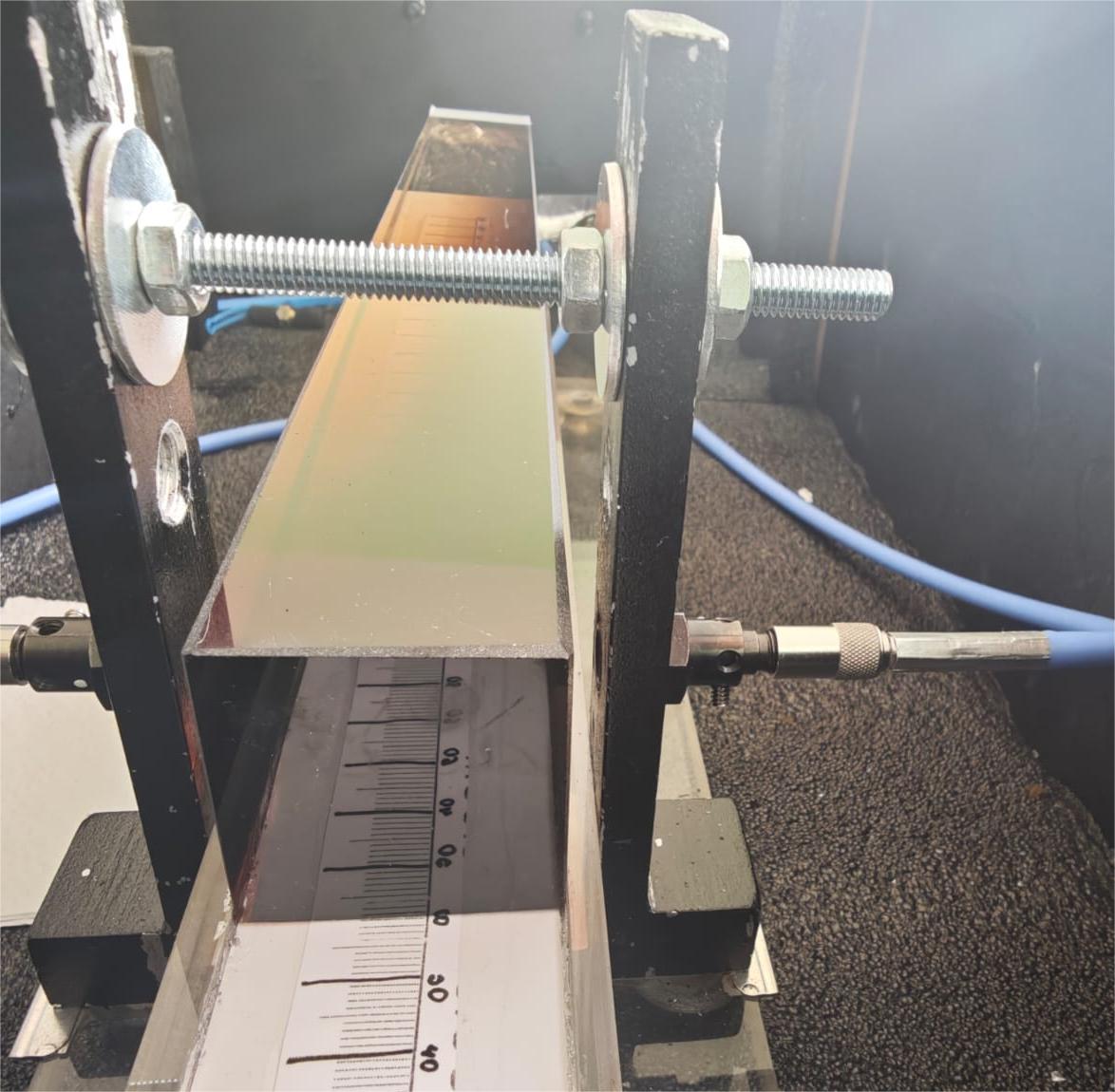}
\includegraphics[width=0.35\linewidth,height=7.0cm,angle=0.0]{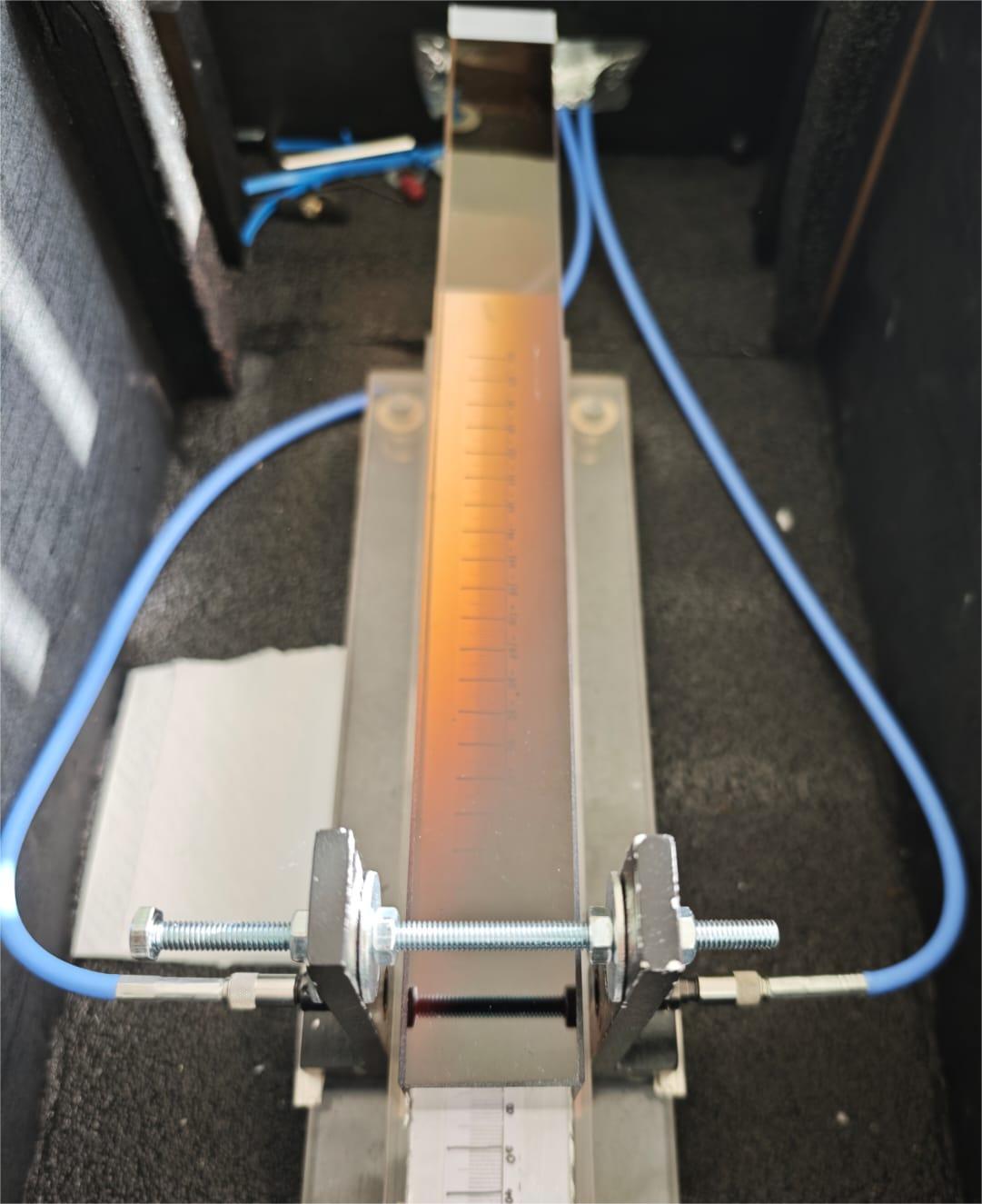}
    \caption{The TF1 lead glass block installed between two fibers for optical transmittance measurement after thermal annealing.     } 
    \label{fig:TF1-transmit-setup}
    \end{figure}

To quantitatively estimate the annealing effect, optical transmittance measurements were performed on the recovered crystals. 
As shown in Figure~\ref{fig:TF1-transmit-setup}, the TF-1 crystal was installed between two optical fibers for transmittance measurements.
Light was delivered from a hydrogen-deuteron source through one of the quartz fibers, which passed through the crystal, was recorded by the opposite fiber, and transferred to the spectrophotometer for measurement.

Despite the presence of optical lenses at the front of both fibers, owing to their small diameter ($\sim 50~\mu$m), some of the scattered or deflected light in the crystal may not be recorded in the opposite fiber.
Moreover, the amount of unrecorded light also depends on the thickness of the crystal; thus, it is different for TF-1 and PbWO$_4$.
Consequently, the error in the measured transmittance was mainly due to this effect, which, according to our estimate, was no more than 5-10\%.

\begin{figure}
\centering
\includegraphics[width=0.45\linewidth,height=5.0cm,angle=0.0]{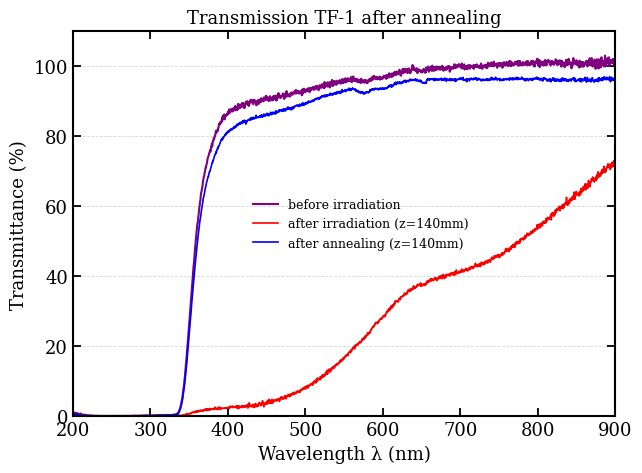}
\includegraphics[width=0.45\linewidth,height=5.0cm,angle=0.0]{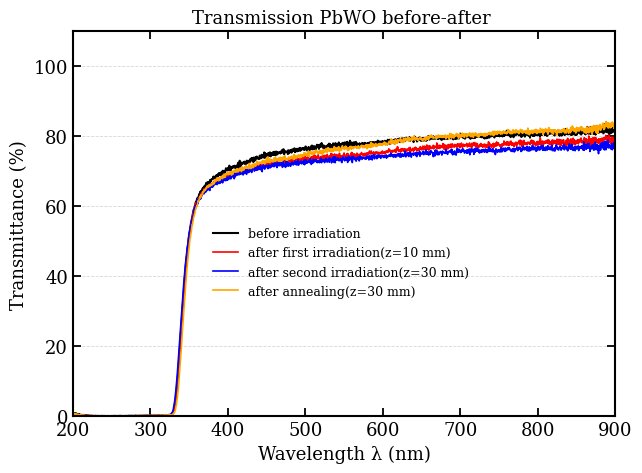}
    \caption{Optical transmittance of lead glass (TF-1) and lead tungstate (PbWO$_4$) crystals before radiation, after irradiation, and after thermal annealing. No corrections for reflections at the crystal surfaces  were applied.} 
    \label{fig:TF1-PbWO-recov}
    \end{figure}

Figure~\ref{fig:TF1-PbWO-recov} shows the optical transmittance of lead glass (TF-1) and lead tungstate (PbWO$_4$) crystals before radiation, after irradiation, and after thermal annealing. No corrections were made for the reflections from the crystal surfaces.
Notably, higher temperature annealing (700-1000$^\circ$C) is more effective and can significantly recover the transmittance of the crystal. 
The transmittance of the annealed crystals increased with increasing annealing temperature and time.


\section{Summary and Outlook}
\label{summary}

Many of the homogeneous electromagnetic calorimeters used or designed for high-energy physics experiments are based on crystals, such as lead glass F8, TF-1, and lead tungstate PbWO$_4$. The latter is the main candidate for the EIC electromagnetic calorimeter. The stability of the crystal performance and their resistance to experimental radiation conditions are very important.
This is the main reason for our study on the radiation damage caused by 20 MeV electrons in lead glass (TF-1) and lead tungstate (PbWO$_4$).

We studied the radiation hardness of two types of crystals, lead glass TF-1 and lead tungstate PbWO$_4$, under 20 MeV energy e-beams from the AANL linear accelerator at LUE-75. 
The optical transmittance of the crystals in the range of wavelength 200-1000 nm was measured before and after irradiation, as well as after thermal annealing. 

The irradiation was performed in two stages. First, both crystals were irradiated with an electron beam current of 0.125 $\mu$A, for a total exposure of 720 s, each absorbing $5.6 \times 10^{14}$ e$^-$. 

Strong degradation of the optical properties of TF-1 caused by this amount of radiation was observed, while the effect in PbWO$_4$ was negligible (it is a few tens of times more radiation hard than TF-1). 
In the second stage, only the PbWO$_4$ crystal was exposed to additional radiation, with a beam current of 0.28 $\mu$A and an exposure time of 1200 s, absorbing additional $2.1 \times 10^{15}$ e$^-$.

Thermal annealing was performed in the temperature range of 160 to 250$^\circ$C (isochronal) for 10-12 hours. The transmittance of the annealed crystals increased with increasing annealing temperature and time.
Note that low-temperature (200-300$^\circ$C) thermal annealing does not change the crystal composition, but it can significantly reduce radiation-induced damage (i.e., color centers) and generate structural relaxation.

This is the first low-energy radiation hardness estimation for a new type of PbWO$_4$  (PWO-II-type) crystal produced by Crytur, a potential provider of the EmCal calorimeter of the ePIC detector at the future EIC. 
The main error in the absorbed total radiation dose in our measurements was due to the low-energy electromagnetic background from the environment surrounding the beam core and the precise determination of the number of electrons absorbed by the crystal. 
Therefore, we could not accurately determine the total number of electrons absorbed by the catalyst. However, according to our estimates, the error should not exceed 15-20\%.


\section*{Acknowledgments}

We are grateful to the AANL LUE-75 accelerator staff for creating the electron beam and conditions necessary for these studies. We thank AANL Chief Engineer M. Martirosyan and Radiation Department Operator A. Margaryan for ensuring the radiation safety during this study.
The work of the AANL group was supported by the RA High Education and Science Committee within the framework of Research Project 21AG-1C028.



\end{document}